\documentclass[twocolumn]{aastex631}
\DeclareMathAlphabet{\pazocal}{OMS}{zplm}{m}{n}
\newcommand{\La}{\mathcal{L}}
\newcommand{\Lb}{\pazocal{L}}

\usepackage[utf8]{inputenc} 
\usepackage[T1]{fontenc}    
\usepackage{hyperref}       
\usepackage{url}            
\usepackage{amsfonts}       
\usepackage{xcolor}
\usepackage{graphicx} 
\graphicspath{ {./images/} }
\usepackage{amsmath}
\usepackage{natbib}

\begin{document}

\title[He~{\small{II}} temperature fluctuations]{A New Approach for Constraining Large-Scale Temperature Fluctuations in the Intergalactic Medium}


\author[0000-0001-7542-8915]{Saba Etezad-Razavi}
\affiliation{Department of Astronomy and Astrophysics, The University of Chicago, 5640 S. Ellis Ave., Chicago, IL 60637, USA}
\affiliation{Institute for Theoretical Physics, Heidelberg University, Philosophenweg 12, D–69120, Heidelberg, Germany}
\affiliation{Department of Physics, Sharif University of Technology, Tehran 11155-9161, Iran}
\affiliation{Max-Planck-Institut für Astronomie, Königstuhl 17, 69117 Heidelberg, Germany}
\affiliation{Perimeter Institute for Theoretical Physics, N2L 2Y5 Waterloo, Canada}
\affiliation{University of Waterloo,  200 University Ave W, Waterloo, ON N2L 3G1, Canada}

\author[0000-0001-8582-7012]{Sarah E.~I.~Bosman}
\affiliation{Institute for Theoretical Physics, Heidelberg University, Philosophenweg 12, D–69120, Heidelberg, Germany}
\affiliation{Max-Planck-Institut für Astronomie, Königstuhl 17, 69117 Heidelberg, Germany}

\author[0000-0003-0821-3644]{Frederick B.~Davies}
\affiliation{Max-Planck-Institut für Astronomie, Königstuhl 17, 69117 Heidelberg, Germany}





\begin{abstract}
The reionization of helium is thought to occur at $2.5\lesssim z\lesssim4$, marking the last phase transition and final global heating event of the intergalactic medium (IGM). Since it is driven by rare quasars, helium reionization should give rise to strong temperature fluctuations in the IGM between neutral and recently-ionized regions of order $\sigma (\ln T) \sim \Delta T/T = 20-50\%$. We introduce a novel method to search for reionization-induced temperature fluctuations in the IGM by using the effective optical depths of the Lyman-$\alpha$ forest towards a large number of background quasars. 
Higher IGM temperatures give rise to lower effective optical depths in the Lyman-$\alpha$ forest, implying that temperature fluctuations will broaden the observed optical depth distribution. 
We measured the distributions of effective Lyman-$\alpha$ forest optical depths across  $71$ X-Shooter spectra from the XQ-100 survey in four redshift bins from $z=3.76$ to $z=4.19$ and compared them to a large-volume cosmological hydrodynamical simulation.
A good agreement is found between the observations and the simulation, which does not include temperature fluctuations; therefore, we do not detect a signature of helium reionization. 
We then post-process the simulations to include an increasing amount of temperature fluctuations until the model becomes inconsistent with the observations. We obtain tight constraints on $\sigma (\ln T) < 0.29 \ (<0.40)$ at $2 \sigma\ (3 \sigma)$ at $z=3.76$ when averaging over scales of $100$ comoving Mpc, and weaker constraints for higher redshifts and smaller scales. 
Our constraints are the tightest to date, and imply that either the IGM temperature contrast caused by helium reionization is less than $\sim30\%$, or that the process has not yet significantly started at $z=3.76$.
\end{abstract}

\keywords{Intergalactic medium - Reionization - Lyman alpha forest - Large-scale structure of the universe 
}

\section{Introduction}\label{intro}

Helium reionization marks the final phase transition of the diffuse baryonic matter which makes up the Inter-Galactic Medium (IGM). While the reionization of hydrogen is thought to be driven primarily by emission from galaxies and to finish by $z\sim5.3$ (e.g.~\citealt{Robertson15, Bosman22}), helium reionization is limited by the availability of high-energy photons ($E>54.5$ eV) coming from luminous quasars \citep{Madau94,Miralda00,McQuinn09, Compostella13,Compostella14}. Calculations of the ionizing photon output from the known abundance of quasars indicate that the helium reionization process is expected to end roughly at $z\sim3$ and last for approximately $1$ Gyr \citep{haardt_madau, Laplante16, Khaire17, Kulkarni19Q,Worseck19}.  

During reionization processes, the excess energy deposited into the photoionized electrons is redistributed into the IGM, causing a global increase in the IGM temperature (e.g.~\citealt{Miralda94}). While the exact degree of heat injection by helium reionization is somewhat uncertain due to the unknown photon spectral index at $E>54.4$\,eV (e.g.~\citealt{UptonSanderbeck16}),
it is expected to increase the local IGM temperature by a theoretical range of $\Delta T\sim0.5-3\times10^4$\,K \citep{1999ApJ...523...66A}, set by the optically-thin and optically-thick limits of ionizing photon absorption. More detailed calculations suggest a range from $\Delta T\simeq1.5\times10^4$\,K (e.g.~\citealt{2008ApJ...682...14F,McQuinn09}) down to $\Delta T\simeq0.8\times10^4$\,K \citep{UptonSanderbeck16}, where the latter is more consistent with the ``bump'' in the IGM thermal history (e.g.~\citealt{Gaikwad21}). After this heating event the IGM then slowly cools over the next few Gyr before settling back down to the pseudo-equilibrium state set (primarily) by the competition between the photoheating by the metagalactic ionizing background and Compton cooling by CMB photons \citep{McQuinn16}.

\begin{figure*}[t]
\centering
\includegraphics[scale=.45]{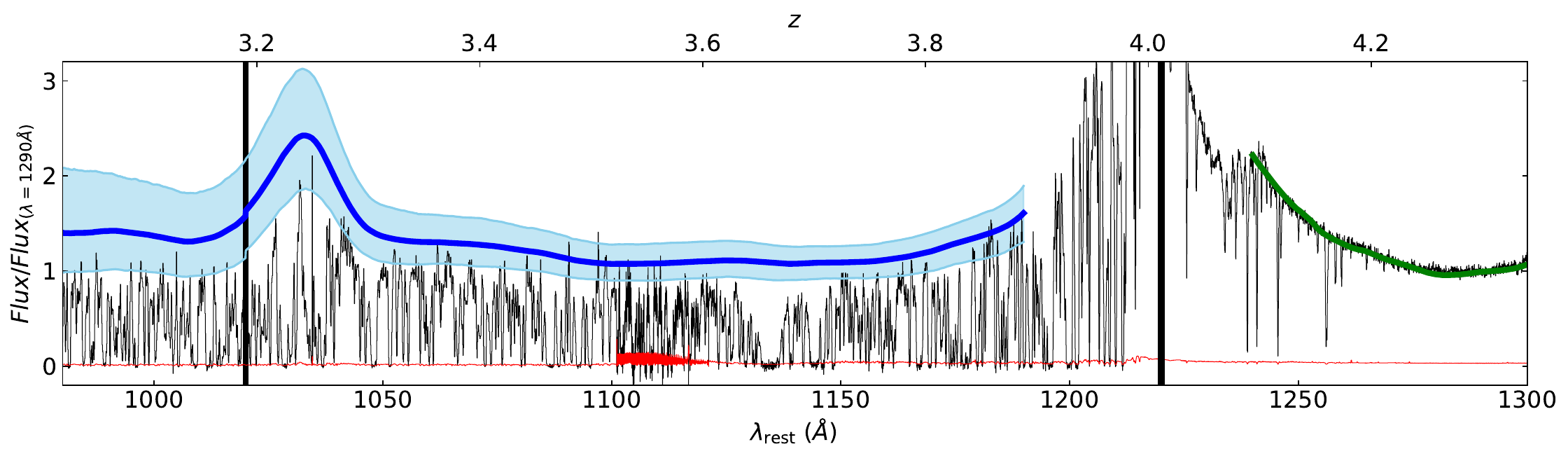}
\caption{Illustrative X-Shooter quasar spectrum from XQ-100: J2215$-$1611 at $z=3.995$, shown in the rest frame of the quasar. Black shows the flux normalized at wavelength $\lambda=1290$\AA, and red shows its uncertainty. Black vertical lines indicate the wavelengths of Lyman-$\beta$ and Lyman-$\alpha$. The solid blue line corresponds to our PCA prediction of the quasar's underlying continuum, with the blue-shaded region showing the $1\sigma$ uncertainty. We employ wavelengths $1060 <\lambda_{\rm{rest}}< 1185$ \AA \ in our measurements of the optical depth.}
\label{fig:example}
\end{figure*}

At $z\sim3$--$4$, regions of the IGM which have not yet undergone helium reionization will have typical temperatures of $\sim7000\,$K \citep{McQuinn16}. Given the range of possible heat injections discussed above, there will thus be a contrast of a factor of two to three in temperature between recently-reionized (hot) regions and not-yet-reionized (cold) regions. Due to the expected large-scale coherence of the helium reionization topology from the rarity of the bright quasars that drive the process (e.g.~\citealt{McQuinn09}), it is possible that the resulting temperature fluctuations impart detectable signatures in the statistics of the (hydrogen) Ly$\alpha$ forest \citep{Lai_2006}. Such a detection would provide valuable constraints on the timing and topology of the helium reionization process. However, the impact of these temperature fluctuations on the 1D (line-of-sight) Ly$\alpha$ forest power spectra is expected to be very weak, on the order of $5\%$ \citep{McQuinn11}, with a stronger signal on the order of $\sim25\%$ potentially visible in the 3D power spectrum \citep{McQuinn11,Greig15}.

Here we explore the constraining power of an alternative, and much simpler, statistic: the distribution of large-scale effective optical depths of the Ly$\alpha$ forest. This distribution has historically been used to constrain the end stages of the \emph{hydrogen} reionization process at $z\sim5$--$6$, where the last remaining neutral islands imprint large-scale Gunn-Peterson troughs in the Ly$\alpha$ forest that significantly broaden the distribution relative to the post-reionization expectation from the density field alone \citep{Becker15,D'Aloisio15,Davies16,Kulkarni19,Nasir20}. Recently, \citet{Bosman22} showed that the \emph{lack} of excess Ly$\alpha$ forest fluctuations at lower redshifts, as quantified by the effective optical depth distribution in the XQR-30 sample of high-redshift quasar spectra \citep{D'Odorico23}, could be used to pinpoint the end of the hydrogen reionization process. Motivated by their success, we perform a similar analysis at $z\sim4$, comparing spectroscopic quasar observations from the XQ-100 survey \citep{xshooter2016} to predictions from Nyx cosmological hydrodynamical simulations \citep{Almgren_2013,Lukic15}. 




We adopt cosmological parameters from \citet{planck}, with $H_0=67.4\  \rm{km}\ \rm{s}^{-1}\ \rm{Mpc}^{-1}$ and $\Omega_m = 0.315$.
The paper's organization is as follows: in Section \ref{data}, we start by describing the XQ-100 data.
In Section~\ref{method}, we explain our methodology including our quasar continuum-fitting procedure using principal component analysis (PCA) in Section~\ref{PCA}. Details on our masking routine and DLA exclusion are provided in \ref{masking} and \ref{dla}. The measurements of the effective optical depth in multiple redshift bins is explained in Section~\ref{optical}. 
We describe our models and the simulations that we use in Section~\ref{model}, and the forward-modeling of the simulations with the procedure for adding temperature fluctuations to our sightlines is detailed in Section~\ref{forward}. Finally, we describe our statistical inference procedure in Section~\ref{likelihood}. 
We present our results from the cumulative distribution functions and constraints on temperature fluctuations in Section~\ref{results} and a discussion of the implications of our measurements for existing He~{\small{II}} reionization models in Section~\ref{discussion}. We finish with a conclusion and summary in Section~\ref{conslusion}.

\section{Data} \label{data}

We use the Lyman-$\alpha$ forest in the spectrum of the quasars to measure the amount of absorption in the IGM due to diffuse gas along the line of sight to the quasars at different redshifts. For this, we need a sample of high signal-to-noise ratio (SNR) quasar spectra covering the wavelength range $1026<\lambda_{\rm{rest}} <1190$ \AA \ to capture the properties of the IGM.

We used the XQ-100 legacy survey \citep{xshooter2016} consisting of $100$ quasars at redshift $3.5<z<4.5$ observed with the X-Shooter spectrograph on the Very Large Telescope (VLT). The X-Shooter spectrograph is the first of the second-generation instruments of the VLT \citep{refId0}. X-Shooter consists of three arms, the UV-Blue arm ($3150 - 5600$ \AA), Visible ($5400 - 10200$ \AA), and Near-IR ($10000 - 24800$ \AA) arms. 
We used the publicly-released reductions of the spectra.

The XQ-100 survey observed with full spectral coverage from $3150$ to $25000$ \AA \ at a resolving power ranging 
from $R \sim 4000$ to $7000$, depending on wavelength. The exposure time along each arm is, $T_{\rm{exposure}} = 890 \rm{s}$ in UVB,
$T_{\rm{exposure}} = 840 \rm{s}$ in VIS and $T_{\rm{exposure}} = 900 \rm{s}$ in the NIR.
The median SNR are $33$, $25$ and $43$, as measured at rest-frame wavelengths $1700$, $3000$ and $3600$ \AA, respectively \citep{xshooter2016}. The angular distribution of the XQ-100 quasars is over the full sky, 
having only two quasars closer than $1^{\circ}$ to each other. 



The unabsorbed quasar continua, on which we base the absorbed continua reconstructions, fall within the VIS arm, while IGM absorption falls in the VIS arm ($z\gtrsim3.7$) or the UV arm ($z\lesssim3.6$). In order to measure the optical depths at $z<3.6$, we, therefore, attempted to stitch the VIS and UV spectra by rescaling the UV spectra to match the flux in the overlapping spectral range ($5400$\AA$ <\lambda< 5600 $\AA). 
We tried to validate the accuracy of the stitching procedure by analyzing the flux ratio of spectra of the same quasars observed in XQ-100 and by 
the SDSS-IV Extended BOSS \citep{eboss} spectrograph (which require no stitching). Our test revealed large and variable errors of order $\sim20\%$ in the fluxing of the UV arm of X-Shooter (see Appendix~\ref{appendix-bin}). 
This is likely caused by a known issue with the X-Shooter VIS arm whereby the response at the edge of the first VIS order is occasionally seen to drop by a large fraction for reasons which are not fully understood (c.f.~section 4.2 of \citealt{Verro22}).
Since the XQ-100 spectra are the product of many co-added observed frames, the resulting error is complex and resolving it is beyond the scope of this work. 

Not all XQ-100 quasars were observed as part of eBOSS to enable such a comparison. Still, for those which were, we noticed some broad-line variability between the eBOSS and XQ-100 spectra (generally taken at a later time). We, therefore, cannot 
use continuum reconstructions performed using the high-SNR X-Shooter VIS observations to analyze the eBOSS spectra at $\lambda < 5600$ \AA \ since the broad lines may (and in some cases did) vary. 

This X-Shooter issue leads us to exclude the UV arm from our analysis for the time being. Namely, we are excluding all wavelengths, $\lambda < 5600$ \AA \ in the observed frame. 
In reality, the flux calibration issue at the edge of the bluest order of the VIS arm begins before the stitching point with the UV arm; we use the ratio of spectra in of the quasars in XQ-100 and eBOSS to pinpoint the range of observed wavelengths which need to be excluded. We find that the deviation begins at a wavelength corresponding to Lyman-$\alpha$ at $z=3.68$ and use this to define our redshift bins starting at this point, in consecutive intervals 100 cMpc in length (see Appendix~\ref{appendix-bin}).

Additionally, we do not use quasars J0747+2739 and J1108+1209, as our PCA method fails dramatically in fitting the quasar continuum in these cases (see Section \ref{PCA} for the PCA method). This is probably due to the lack of anything similar to those objects in the PCA's training set; potentially, these quasars are weak broad absorption line (BAL) quasars or are otherwise anomalous. After the exclusion of these quasars and the redshift constraints described in the previous paragraph, we are left with $71$ usable quasars from XQ-100 which are listed in Table \ref{qso1}.



\begin{table}
\centering
\begin{tabular}{  c |  c | c | c }
\cline{1-4}
{XQ-100 name} & {$z_{\rm{qso}}$}  & \textbf{ ${\rm{SNR}}_{1700 {\text{\AA}}}$}  & {Comments} \\
\cline{1-4}
J0003$-$2603 & 4.125 &79 & 1 $\text{DLA}_{3000}$ , 1 $\text{DLA}_{5000}$  \\
J0006$-$6208 & 4.440 & 20 & 2 $\text{DLA}_{5000}$\\
J0030$-$5129 & 4.173 & 18 & -\\
J0034+1639 & 4.292 & 28 & 1 $\text{DLA}_{3000}$, 3 $\text{DLA}_{5000}$\\
J0042$-$1020 & 3.863 & 52 & 1 $\text{DLA}_{5000}$\\
J0048$-$2442 & 4.083 & 20 & 1 $\text{DLA}_{3000}$\\
J0113$-$2803 & 4.314 & 30 & 1 $\text{DLA}_{3000}$, 1 $\text{DLA}_{5000}$\\
J0117+1552 & 4.243 & 40 & - \\
J0121+0347 & 4.125 & 31 & 1 $\text{DLA}_{3000}$\\
J0124+0044 & 3.837 & 34 & 1 $\text{DLA}_{3000}$\\
J0132+1341 & 4.152 & 32 & 1 $\text{DLA}_{3000}$\\
J0133+0400 & 4.185 & 48 & 2 $\text{DLA}_{3000}$, 2 $\text{DLA}_{5000}$\\
J0137$-$4224 & 3.971 &  17 & 2 $\text{DLA}_{3000}$\\
J0153$-$0011 & 4.195 & 15 & 1 $\text{DLA}_{3000}$\\
J0211+1107 & 3.973 & 22 & 2 $\text{DLA}_{3000}$\\
J0214$-$0517 & 3.977 & 31 & 1 $\text{DLA}_{5000}$\\
J0234$-$1806 & 4.305 & 28 & 1 $\text{DLA}_{3000}$, 1 $\text{DLA}_{5000}$\\
J0244$-$0134 & 4.055 & 39 & 2 $\text{DLA}_{3000}$\\
J0247$-$0556 & 4.234 & 22 & 1 $\text{DLA}_{3000}$\\
J0248+1802 & 4.439 & 26 & - \\
J0255+0048 & 4.003 & 30 & 1 $\text{DLA}_{3000}$, 2 $\text{DLA}_{5000}$\\
J0307$-$4945 & 4.716 & 37 & 1 $\text{DLA}_{3000}$, 2 $\text{DLA}_{5000}$\\
J0311$-$1722 & 4.034 & 39 & 1 $\text{DLA}_{3000}$\\
J0403$-$1703 & 4.227 & 21 & 1 $\text{DLA}_{3000}$\\
J0415$-$4357 & 4.073 & 16 & 1 $\text{DLA}_{3000}$, 1 $\text{DLA}_{5000}$\\
J0426$-$2202 & 4.329 & 26 & 1 $\text{DLA}_{5000}$\\
J0525$-$3343 & 4.385 & 39 & - \\
J0529$-$3526 & 4.418 & 22 & 1 $\text{DLA}_{3000}$\\
J0529$-$3552 & 4.172 & 13 & 2 $\text{DLA}_{3000}$\\
J0714$-$6455 & 4.465 & 29 & - \\
J0800+1920 & 3.948 & 29 & 2 $\text{DLA}_{3000}$\\
J0833+0959 & 3.716 & 33 & -\\
J0835+0650 & 4.007 & 33 & 2 $\text{DLA}_{3000}$, 1 $\text{DLA}_{5000}$\\
J0839+0318 & 4.230 & 12 & 1 $\text{DLA}_{3000}$\\
J0935+0022 & 3.747 & 27 & -\\
J0937+0828 & 3.704 & 23 & - \\
J0955$-$0130 & 4.418 & 35 & 1 $\text{DLA}_{3000}$, 1 $\text{DLA}_{5000}$\\
J0959+1312 & 4.092 & 54 & 1 $\text{DLA}_{3000}$\\
J1013+0650 &  3.809 & 30 & 1 $\text{DLA}_{3000}$\\
J1032+0927 & 3.985 & 27 & 1 $\text{DLA}_{3000}$\\
J1034+1102 & 4.269 & 33 & -\\
J1036$-$0343 & 4.531 & 19 & 1 $\text{DLA}_{3000}$\\
J1037+0704 & 4.127 & 52 & 1 $\text{DLA}_{3000}$\\
J1054+0215 & 3.971 & 14 & -\\
J1057+1910 & 4.128 & 19 & 3 $\text{DLA}_{3000}$\\
\cline{1-4}
\end{tabular}
\caption{XQ-100 quasars used in this work. The first column gives the XQ-100 name of each quasar, with a systemic redshift provided in the next column. The third column lists the signal-to-noise ratio for each quasar.} 
\label{qso1}
\end{table}

\begin{table}
\centering
\begin{tabular}{  c |  c | c | c }
\cline{1-4}
{XQ-100 name} & {$z_{\rm{qso}}$}  & \textbf{ ${\rm{SNR}}_{1700 {\text{\AA}}}$}  & {Comments} \\
\cline{1-4}
J1058+1245 & 4.341 & 26 & BAL, 1 $\text{DLA}_{5000}$ \\
J1110+0244 & 4.146 & 30 & -\\
J1111$-$0804 & 3.922 & 43 & 1 $\text{DLA}_{3000}$, 1 $\text{DLA}_{5000}$\\
J1126$-$0124 & 3.765 & 22 & -\\
J1135+0842 & 3.834 & 55 & -\\
J1248+1304 & 3.721 & 39 & 1 $\text{DLA}_{3000}$\\

J1312+0841 & 3.731 & 33 & 1 $\text{DLA}_{5000}$\\
J1320$-$0523 & 3.717 & 41 & 1 $\text{DLA}_{3000}$\\
J1323+1405 & 4.054 & 23 & -\\
J1330$-$2522 & 3.949 & 39 & 1 $\text{DLA}_{3000}$\\
J1331+1015 & 3.852 & 33 & -\\
J1336+0243 & 3.801 & 33 & 1 $\text{DLA}_{3000}$\\
J1352+1303 & 3.706 & 14 & -\\
J1401+0244 & 4.408 & 39 & -\\
J1542+0955 & 3.986 & 31 & 1 $\text{DLA}_{3000}$\\

J1552+1005 & 3.722 & 35 & 1 $\text{DLA}_{3000}$, 2 $\text{DLA}_{5000}$\\
J1621$-$0042 & 3.711 & 34 & 1 $\text{DLA}_{3000}$\\
J1633+1411 & 4.365 & 31 & -\\
J1658$-$0739 & 3.750 & 37 & 2 $\text{DLA}_{3000}$\\
J1723+2243 & 4.531 & 16 & 1 $\text{DLA}_{5000}$\\
J2215$-$1611 & 3.995 & 40 & 5 $\text{DLA}_{3000}$\\
J2216$-$6714 & 4.479 & 21 & 1 $\text{DLA}_{3000}$\\
J2239$-$0552 & 4.557 & 10 & 1 $\text{DLA}_{5000}$\\
J2251$-$1227 & 4.157 & 34 & 2 $\text{DLA}_{3000}$\\
J2344+0342 & 4.248 & 32 & 1 $\text{DLA}_{3000}$, 1 $\text{DLA}_{5000}$\\
J2349$-$3712 & 4.219 & 21 & 2 $\text{DLA}_{3000}$\\
\cline{1-4}
\end{tabular}
\vskip 0.5em
\normalsize{Table 1, continued}
\end{table}

\section{Methods}\label{method}
We study the evolution of IGM transmission between redshifts $3.7<z<4.2$, with the aim of constraining the temperature fluctuations in IGM during the He~{\small{II}} reionization. For this goal, we use $71$ high SNR quasar spectra and we re-construct the quasar's intrinsic emitted continua in the Lyman-$\alpha$ forest using PCA. Having the quasars' intrinsic continua from the PCA reconstruction and the observed spectra, we measure the transmission and optical depth in the Lyman-$\alpha$ forest of the quasars. In the last step, we compare our observations of effective optical depth to simulations to constrain temperature fluctuations in the IGM. In this Section, we first describe our continuum fitting, PCA techniques and optical depth measurements in the subsections \ref{PCA} to \ref{optical}. We then discuss our model and simulations in subsection \ref{model}, and finally, we present our likelihood measurement in subsection \ref{likelihood}.


\subsection{PCA to reconstruct the underlying continum}\label{PCA}

We employ PCA to reconstruct the quasars' intrinsic continua, 
$F_{\rm{cont}}$ at $\lambda_{\rm{rest}} < 1190$\AA \ using the observed quasar continuum at $\lambda_{\rm{rest}} > 1280$\AA.  Here, we use a PCA method developed by \citet{bosman_2021}, based on the log-PCA approach of \citet{davies_2018c} (see also \citealt{davies_2018b}) and further refined as described in \citet{Bosman22}. This PCA method achieves the best current accuracy in the reconstruction of the continuum in the Lyman-$\alpha$ forest. We briefly summarize the method below. For a more detailed discussion of the training and testing schemes of our PCA and a comparison to other methods, we refer the reader to \citet{bosman_2021}.

The PCA is constructed by obtained by using a training set of spectra of low-redshift quasars to find optimal linear decompositions of the `known' red side ($\lambda> 1280$\AA) and the `unknown' blue side of the spectrum ($\lambda < 1220$\AA). An optimal mapping is then determined between the linear coefficients of the two sides' decompositions \citep{1993AJ....106..417F,yip_2004, Suzuki06, paris_2011, duro}. 
The training set includes $4597$ quasars at $2.7 < z < 3.5$ with SNR $>  7$ from the SDSS-III Baryon Oscillation Spectroscopic Survey \citep{boss} and eBOSS. 
The reconstruction uncertainty of the PCA method, after testing on an independent set of $4597$ quasars from eBOSS, is ${\rm{PCA}}/{\rm{True}} -1 = 0.8^{+7.8}_{-7.9}\%$, i.e.~the method predicts the underlying continuum within $8\%$ with a negligible bias and a weak wavelength dependence. The asymmetric $1\sigma$ and $2\sigma$ bounds are measured empirically by finding the central $68$th and $95$th percentile intervals of the prediction error in the testing sample.
 
Our PCA consists of $15$ red-side components and $10$ blue-side components that are used to fit the red-side continuum and to reconstruct the blue-side continuum. 
For the fitting to the red-side continuum, we first automatically fit a slow-varying spline to which the PCA components are then fitted. The auto-spline continuum fitting we are using \citep{davies_2018c} is based on a modified version of the method of \citet{dallaglio}, based initially on the procedures outlined in \citet{spline_young}, and \citet{carswell_1991}. This step is done to make the PCA less biased by random noise in the spectra. 

. 

\begin{table}
\centering
\begin{tabular}{  c |  c | c }
\cline{1-3}
{Central $z$} & { $N_{\rm{los}}$ with $L = 100$ Mpc}  & {$N_{\rm{los}}$ with $L = 50$ Mpc} \\
\cline{1-3}
3.76 & 45 & 45 \\
3.90 & 36 & 35 \\
4.04 & 23 & 24 \\
4.19 & 13 & 13  \\
\cline{1-3}
\end{tabular}
\caption{Number of lines of sight $N_{\rm{los}}$ contributing to the redshift bins with different comoving sizes.}
\label{bins}
\end{table}

\subsection{Masking} \label{masking}

Before conducting our measurements and continuum fitting, we need to mask the regions in the spectrum with instrumental defects and those affected by other physics intervening along the quasars' sightlines. These effects do not represent the underlying quasar emission and thus need to be masked so that our PCA can find the correct fit. To do this we develop an auto-masking procedure which we outline here. 

First, we mask the atmosphere telluric absorption; namely, we mask wavelenghts with high atmospheric absorption, $13450$ \AA$ < \lambda_{\rm{obs}} < 14250$ \AA \ and $18000$ \AA$ < \lambda_{\rm{obs}} < 19450$ \AA.
Second, we mask the regions with high instrumental uncertainties. We apply a sigma-clipping criterion to mask bad pixels; more specifically, any region that, after fitting the auto-spline to the atmosphere-masked flux, has  $|\rm{flux} - \rm{auto\ spline\ fit}| > 5 \sigma$. 
Third, we mask pixels with exceptionally large normalized flux values that can happen due to cosmic rays or residuals in telluric correction, i.e.~pixels with flux values larger than six times that of the Lyman-$\alpha$ emission peak are masked. 

Finally, we  address the potential presence of Broad Absorption Lines (BAL) quasars in our sample. BALs are absorption features created by accelerated gas within the quasar itself \citep{BAL_lynds}. They are broad, meaning they can mimic the intrinsic quasar emission, and they can occur at a range of velocities, such that the PCA cannot learn their profiles. It is therefore necessary to address them manually. We visually identified one broad absorption line (BAL) in our quasar set, $J1058+1245$, for which we manually masked the wavelength region $1420 < \lambda_{\rm{rest}} < 1490$\AA.



\subsection{Exclusion of DLAs}
\label{dla}

To measure the amount of transmission in the IGM, we wish to exclude from our analysis all absorption associated with galaxies intervening along the sightline since our focus is the IGM absorption. Damped Lyman-$\alpha$ (DLA) absorption systems are broad absorption lines in the Lyman-$\alpha$ forest of the quasar, which occur from the concentrations of neutral hydrogen gas associated with galaxies along the line of sight to the quasars \citep{bookDLA}. Due to the lack of an efficient way to simulate the high density systems which result in the observed DLAs, we mask these objects. The existence of these DLA objects does not affect our PCA or auto-spline fitting routines as both of these procedures are applied on the red side of the Lyman-$\alpha$ line, but without masking, it will affect the transmission that we want to measure on the blue side.
We mask DLAs using the DLA catalog by \citet{Berg_2016} with a slight modification. We accept the DLA classification in the catalog only if one of the following conditions is met: if $\log(N_{\rm{HI}}) > 20.3$ cm$^{-2}$, we mask the area around the center of the DLA across a window $\Delta v = 5000$ km s$^{-1}$ ($\text{DLA}_{5000}$ in Table \ref{qso1}); if  $19.0 <\log(N_{\rm{HI}}) < 20.3$ cm$^{-2}$ and corresponding metal absorption is detected (refer to table 4 in \citealt{Berg_2016}), we mask the area around the absorption redshift of these objects with $\Delta v = 3000$ km s$^{-1}$ ($\text{DLA}_{3000}$ in Table \ref{qso1}). 

In addition to DLAs, we still need to mask some other regions on the blue side, $\lambda_{\rm{rest}}<1215.67$\AA. For our studies, we are mainly interested in the state of the IGM on average; hence we are not considering the quasar's proximity zone as it is mainly ionized by the UV emission from the quasars and thus biased \citep{Lidz_2007}. In particular, we do not use wavelengths $1190 < \lambda_{\rm{rest}} < 1230 \AA$. For our current analysis, we exclude the Lyman-$\beta$ forest, $\lambda_{\rm{rest}}<1026 \AA$ in the rest-frame. 

At this point, we have the PCA construction on the red-side, $\lambda_{\rm{rest}} > 1230$ \AA, and prediction on the blue-side, $\lambda_{\rm{rest}} < 1190$ \AA. We are ready to measure the amount of transmission in the Lyman-$\alpha$ forest.

\begin{figure}[t]
\centering
{\includegraphics[width=1.0\linewidth]{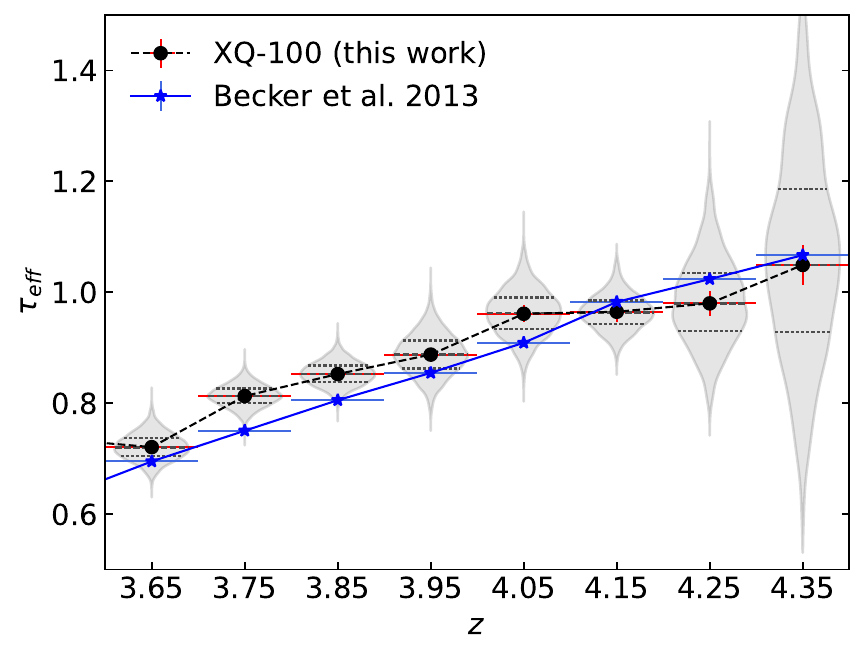}}
\caption{Mean effective optical depths with redshift measured from XQ-100, compared to the literature \citep{Becker_2013}. Uncertainties are obtained from bootstrap resampling; the violins represent the bootstrapped distribution in each redshift bin.  
}
\label{fig:mean}
\end{figure}

\subsection{Optical depth measurements}\label{optical}
We measure the fraction of transmitted flux, $T$, in the Lyman-$\alpha$ forest using the PCA prediction and the flux in all the non-masked pixels as follows:

\begin{equation}
T = \frac{F_{\rm{observed}}}{F_{\rm PCA\ prediction }}.
\end{equation}

We convert the observed wavelenghts inside the Lyman-$\alpha$ forest to the redshift corresponding absorption redshift using $z = \lambda_{\rm{rest}}(z_{\rm{qso}}+1) / \lambda_{{\rm{Ly}}\alpha}  - 1$, where $\lambda_{{\rm{Ly}}\alpha } = 1215.67$ \AA. 
We define bins with equal comoving size $L$. 
Then we divide our data into consecutive bins of that size, starting at $z=3.68$, and stepping up to higher redshifts. 
We stop at $z=4.26$ because less than $10$ sightlines probe higher redshifts. We initially pick a bin size of $100 \rm{Mpc}$ which is potentially the most relevant scale for He~{\small{II}} reionization. 
This gives us the final bin centres that we use in this work: $z=3.76$, $z=3.90$, $=4.04$ and $z=4.19$. 
To probe smaller scales we cut the $100 \rm{Mpc}$ bins into half while keeping the centers of the bins the same as before.

We only use a bin along a specific sightline if at least a comoving length of $L/2$ within the bin is usable, i.e.~un-masked. 
We show the number of sightlines used in each bin in Table \ref{bins}.

Finally, we define the effective optical depth of IGM, $\tau_{\rm{eff}}$, as:

\begin{equation}
\tau_{{\rm{eff}},\ i}  = - \ln \left( \frac{\sum_{j \in I_i} T_j}{N_i} \right) \: 
\end{equation}

where $i$ in the bin number, $I_i$ is the collection of all pixels in the redshift bin $i$ and $N_i$ is the corresponding number of pixels.

Using all redshift bins along our $71$ sightlines, we measure the mean effective optical depth in each redshift bin. We obtain an uncertainity on this effective optical depth using a bootstrap method. The bootstrap distribution is obtained by randomly drawing samples of the same size as the observations in each  redshift bin $200$ times.
Figure~\ref{fig:mean} shows the result of our measurement compared to \citet{Becker_2013}; for the sake of this comparison, we show our measurements in the same linear redshift bins which are used in their work. Error bars demonstrate the standard deviation of the bootstrapped distributions and violins show the complete bootstrap distribution in each redshift bin. The smaller number of sightlines at the higher redshift bins is likely responsible for the non-gaussian shape. However we don't necessarily expect the distributions to be gaussian; this non-gaussianity could be a sign of early reionization around these redshifts. This ``tension'' would get more evident in our likelihood measurements described in Section \ref{likelihood}.
Our sightlines appear to be slightly more absorbed than the means reported in \citet{Becker_2013} based on much larger samples, but this is likely due to statistical chance. Based on the bootstrapped distributions, our mean optical depths measurements agree with \citet{Becker_2013} within $1\sigma$ in all redshift bins except for the one at $z= 3.75$.   

\subsection{Models} \label{model}

%
%
%


We will compare the observed distributions of effective optical depth measured above to those derived from Ly$\alpha$ forest simulations, with and without temperature fluctuations from helium reionization. We describe our modeling procedure below.

\begin{figure*}[t]
\centering
\includegraphics[width=0.8\textwidth]{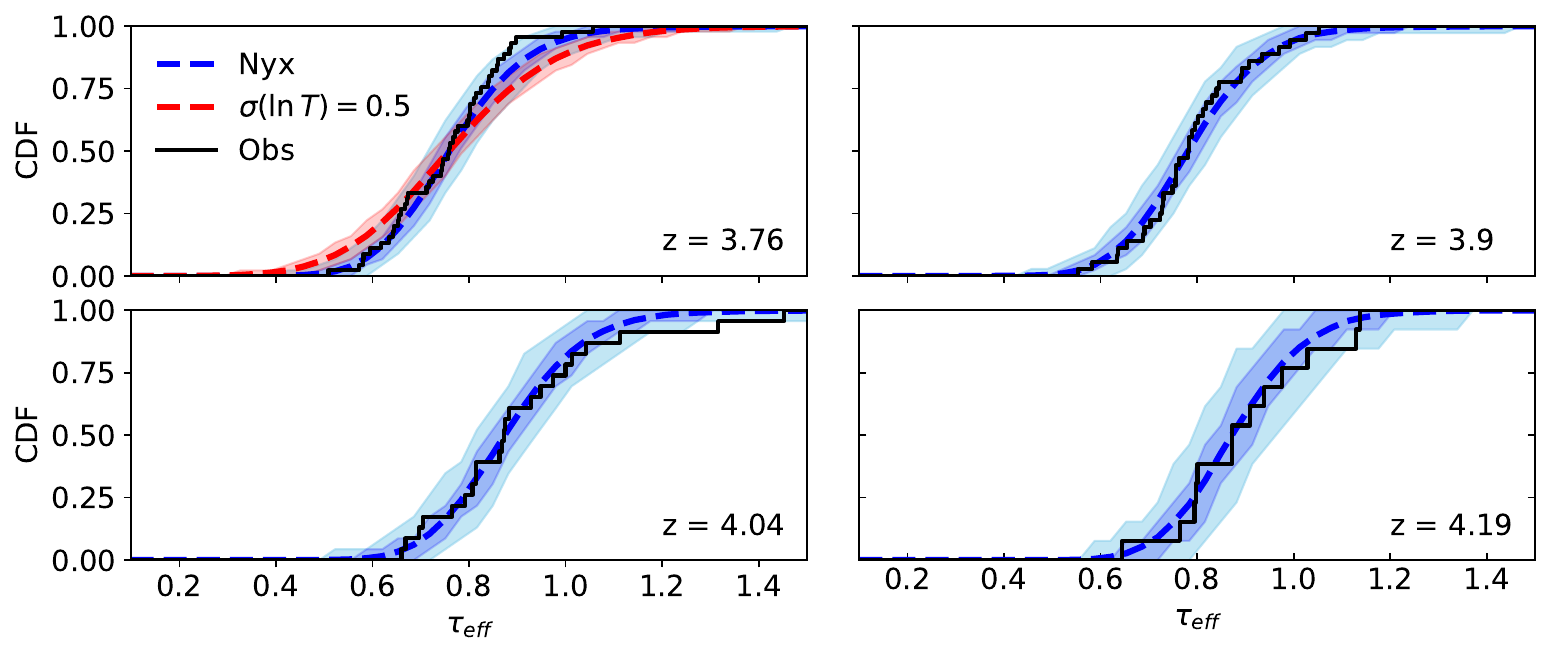}
\caption{Comparison of the Cumulative Distribution Function of effective optical depths between Nyx (blue mean and $1/2\sigma$ contours) and the observations (black line). Statistically, the observations are in agreement with Nyx without any excess fluctuations due to temperature at all redshifts within $2\sigma$.}
\label{2-CDF}
\end{figure*}

We first require simulations of the baseline level of Ly$\alpha$ opacity fluctuations resulting from the density field alone, i.e. the cosmological distribution of matter, on $100 \ \rm{Mpc}$ scales. For this purpose, we post-process snapshots from a cosmological hydrodynamical simulation run with the Nyx code \citep{Almgren_2013} run with a fixed grid of 4096$^3$ baryon cells and the same number of dark matter particles in a volume 100\,Mpc$/h$ on a side. The simulation was run following \citet{Lukic15} with an optically-thin UV background from \citet{haardt_madau}, with snapshots every $\Delta z=0.5$. We extracted 40,000 randomly-oriented skewers of density, temperature, and line-of-sight velocity starting from random locations within the simulation box. We use the snapshot at $z=4$, as it is the closest to the redshift bins of our data, and rescale the physical densities by $(1+z)^3$ to partly correct for this offset. We then compute the neutral hydrogen density along each skewer under the assumption of photoionization equilibrium, and calculate the Ly$\alpha$ opacity including the effects of peculiar motions and thermal broadening (as in, e.g., \citealt{Lukic15} and \citealt{Bosman22}).

We expect helium reionization to imprint large-scale variations in the IGM temperature along the line of sight, but the exact distribution of ionized bubbles and temperature contrast is highly model-dependent. To simplify the interpretation of our measurements, we instead opt for a model in which the average IGM temperature along each observed sightline is drawn from a lognormal distribution of width $\sigma(\ln{T})$. Due to the dependence of the hydrogen recombination rate on temperature, we expect hotter gas to have a lower neutral hydrogen fraction, and vice versa for colder gas. 

While this temperature dependence is analytic for any individual parcel of gas, its effect on the large-scale \emph{effective} optical depth must be calibrated from simulations. We adopt a calibration between IGM temperature and Ly$\alpha$ effective optical depth from \citet{bolton_et_al_2005}, who studied the dependence of the hydrogen photoionization rate inferred from $\tau_{\rm eff}$ as a function of various IGM parameters. From their equation 4
,  and leaving all other parameters fixed, we derive the relationship between $\tau_{\rm eff}$ and $T$ to be $\tau_{\rm eff}\propto T^{0.352}$. We use this expression to map from the lognormal distribution of $T$ fluctuations to additional fluctuations in $\tau_{\rm eff}$.

We note that this model for temperature fluctuations is rather simplistic. In principle one could instead constrain the parameters of a more sophisticated model, e.g. a large-volume simulation of the helium reionization process \citep{McQuinn09,Compostella13,LaPlante17}. We leave a more detailed exploration of helium reionization models to future work.

Next, we will measure the likelihood associated with different amplitudes of the temperature fluctuations, $\sigma(\ln T)$, given our observations, by comparing the modeled and observed distribution of optical depths. 
To achieve this, we must first forward-model the simulation, which we discuss in the following subsection.

\subsubsection{Forward-modeling} \label{forward}

We forward-model the optical depth along the simulated sightlines at each redshift 
to take into account all known sources of uncertainty. The forward-modeling procedure employs the following steps:

\begin{figure}[t]
    \centering
    \includegraphics[width=\linewidth]{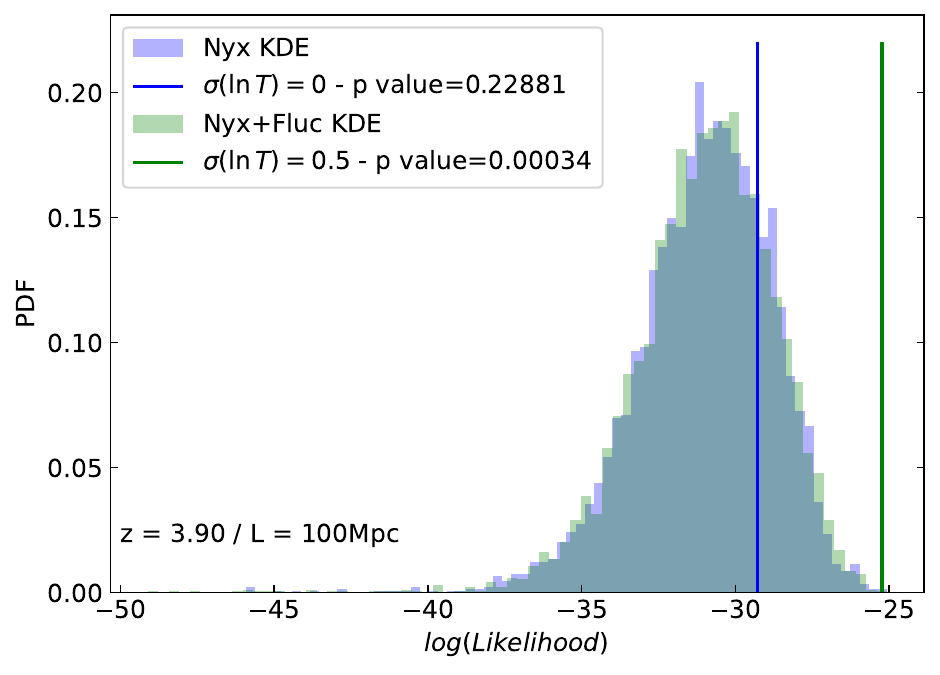}
    \caption{An example of the simulated and observed likelihoods. The solid lines represent the likelihood corresponding to the observed set of optical depths, while the histograms show the likelihood of simulated datasets with the same size and uncertainties as the observations. We measure the distance between the model and observation to obtain the $p$-value of the observations compared to a KDE built from the simulations. By increasing the temperature fluctuations (blue versus green), the distance between the model and observations increases, meaning the probability of occurrence of such an observation given this specific $\sigma(\ln T)=0.5$ decreases.}
    \label{fig:likelihood}
\end{figure}

\begin{itemize}
\item \textbf{Down-sampling:} In the first step, we down-sample our simulated sightline to the XQ-100 resolution at each redshift bin. Simulated sightlines at each redshift of size $100 \rm{Mpc}$ have a number of pixels of $npix \sim 2000$ with slight variations between the 4 different redshift bins. We reduce this to the number of pixels of the observed sightlines in XQ-100 which in each of the redshift bins of size $100 \rm{Mpc}$ is equal to $npix \sim 350$ with slight variation among the bins.
\item \textbf{Adding temperature fluctuations:} We add excess optical depth fluctuations resulting from temperature fluctuations and produced as described in the previous Section, after downsampling the simulated sightlines to the XQ-100 resolution. 
For each sigthline in a redshift bin, we draw a single $\Delta \ln T$ from a gaussian distribution with a standard deviation of $\sigma(\ln T)$. We convert this $\Delta \ln T$ to excess optical depth, $\Delta \ln \tau = -0.352 \Delta \ln T$, and then introduce this optical depth modification via a flat rescaling of the transmitted flux of the entire sightline.


\item \textbf{Adding instrumental noise:} In the third step we add a random error due to instrumental noise (up to $\sim5\%$) to the simulated spectra. We add Gaussian noise to the pixels of each simulated sightline according to the noise vector of the corresponding observed spectrum.


\item \textbf{Adding continuum uncertainity:} We introduce a random shift to the whole continuum, in flux space, due to PCA continuum-reconstruction uncertainty (up to $\sim8\%$). We draw from a normal distribution with the $1\sigma$ width of the PCA uncertainty, as determined from empirical testing on SDSS quasars \citep{bosman_2021}. We add this continuum reconstruction error multiplicatively to all of the pixels of a sightline, i.e.~treating the continuum error as perfectly covariant across the entire spectral segment.

\item \textbf{Mean flux calibration:} To ensure that our simulations are consistent with the observations on average, we rescale the overall optical depths in our simulations by a constant calibration factor, $A$, at each redshift in order to match the observed mean flux: $\langle e^{A\tau_{\rm{sim}}}\rangle = \langle F\rangle_{\rm{obs}}$. This renormalization is equivalent to an adjustment of the assumed ionizing background, which is itself uncertain. The calibration of the simulations has been performed separately in each redshift bin and for all different values of injected excess temperature fluctuations. Namely, to compare each set of observational sightlines in a redshift bin to the simulated ones, we match the mean of the observed flux to the collection of all $\tau_{\rm{eff}}$ in the bin, each for a different sightline, using a different value for $A$, the calibration value.
We checked whether using the median flux value (instead of the mean) for calibration had any impact on our results, as may potentially be the case if the distributions of optical depths are very non-Gaussian. We found that the effect was negligible. 


\end{itemize}


 Figure \ref{2-CDF} shows the cumulative distribution function (CDF) of effective optical depth from our simulations and observations. To estimate the uncertainty we perform a bootstrap resampling by picking a random number $N$ of simulated sightlines at each redshift which is equal to the number of observed sightlines on that bin. Error bars on the simulations are constructed from bootstrap and the solid line shows the mean value of bootstrap/actual measurements. We perform $1000$ bootstrap iterations and estimate the mean and standard deviation of the bootstrap samples at each redshift bin. 
 Figure \ref{2-CDF} shows that our observations are in agreement with the model without any additional temperature fluctuations at all redshifts within $2 \sigma$. This motivates us to use these observations to put an upper limit on the temperature fluctuations. In the next Section we describe our likelihood calculation procedure.
 

\subsection{Likelihood Calculation}\label{likelihood}

\begin{figure}
    \centering
    \includegraphics[width = \linewidth]{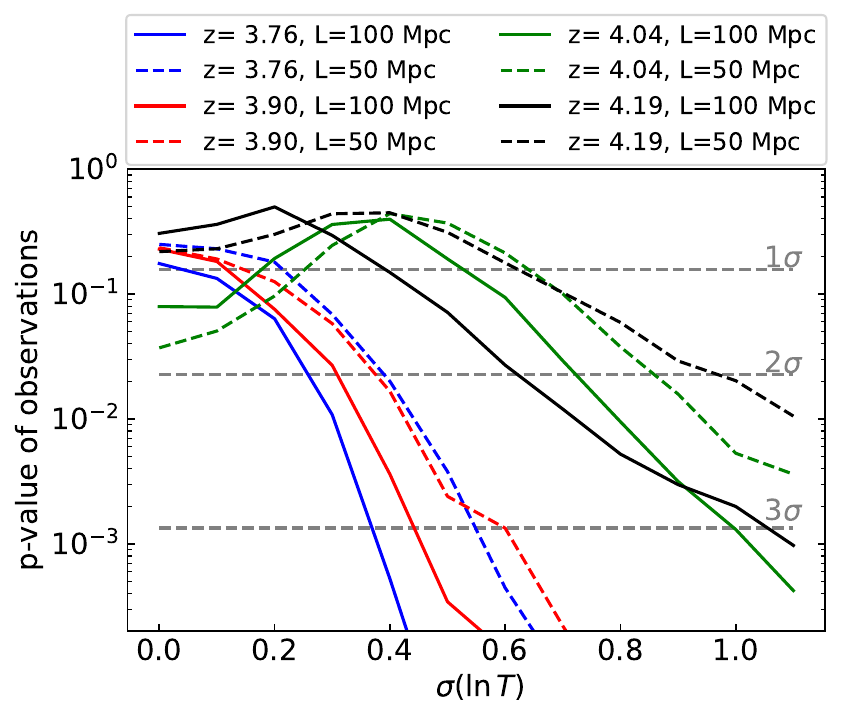}
    \caption{The $p$-value of the observations as a function of the  amount of temperature fluctuations for different redshift bins (colors) and two bin sizes $50\ \rm{Mpc}$ (dashed) and $100\ \rm{Mpc}$ (solid). Blue denotes redshift $z = 3.76$, red $z=3.90$, green $z=4.04$ and black $z=4.19$. Thresholds corresponding to $1,2$ and $3 \sigma$ are indicated.}
    \label{fig:5pvalue}
\end{figure}

Temperature fluctuations from helium reionization could cause extra scatter in the effective optical depths between the sightlines. We quantify the amount of excess temperature fluctuations allowed by our observations with a Bayesian likelihood analysis. Using Bayes theorem assuming a flat prior for the temperature fluctuations, the likelihood of a model with some $\sigma(\ln T)$ is proportional to the probability of occurrence of our observations given this model, $P( \sigma( \log T) | \rm{Observations})$ = $P( \rm{Observations} | \sigma( \log T)) \times \rm{Prior}$. Therefore, maximizing the posterior corresponds to maximizing the likelihood. We measure the probability of occurrence of our optical depth observations along each sightline at each redshift bin and both spatial scales, given our simulations with some amount of additional temperature fluctuation $\sigma (\ln T)$ added on top. This is $\rm{Prob}( \rm{Obs}_n | \rm{Model}_{n;\sigma (\ln T)})$ where $n$ indicates the redshift bin. 
If this probability decreases for increasingly large amounts of $\sigma (\ln T)$, the observed distribution of optical depths will become increasingly unlikely to occur by chance, until it is ruled out for a sufficiently large $\sigma (\ln T)$, as illustrated in the top left panel of Figure~\ref{2-CDF}.

We determine the likelihood of the observations at each redshift by computing the product of the likelihoods of each of the observed $\tau_{\rm{obs,i}}$ in each sightline.
First, for each observed $\tau_{\rm{obs,i}}^n$ in the redshift bin $n$ with uncertainties of $S_{\rm{obs,i}}^n$, we use kernel density estimation (KDE) applied to the post-processed simulated sightlines with the uncertainties corresponding to that observation. For each observation we randomly select $5000$ simulated sightlines and we post-process them using the uncertainties corresponding to that observation and we construct a KDE representation of the distribution of $\tau_{\rm sim, i}$ from these $5000$ simulated sightlines. Using $5000$ sightlines to generate the KDEs corresponds to a constraining power up to $3.5 \sigma$ confidence using the relation ${\rm{stdev}} = \sqrt{2}{\rm{erf}}^{-1}(2p)$ where we let $p = \frac{1}{\sqrt{N}}$ with $N=5000$. We denote the set of all the post-processed simulated sightlines as $\{\tau_{\rm{sim}}^n( S_{\rm{obs,i}}^n)\}$ and build the KDE using them,
\begin{equation}
    \{\tau_{\rm{sim}}^n( S_{\rm{obs,i}}^n)\} \rightarrow \rm{KDE}_i^n.
\end{equation}
This corresponds to a KDE for the observation $i$ in the redshift bin $n$. Then we can estimate the likelihood of occurrence of the observation $\tau_{\rm{obs,i}}^n$ given this KDE as:

\begin{equation}
    \Lb_{{\rm{obs,i}}}^n = \rm{KDE}_i^n(\tau_{\rm{obs,i}}^n),
 \label{eq:likelihood-obsi}
\end{equation}

and finally, to find the likelihood of drawing the full observed dataset in this redshift bin given our model, we combine the likelihood of the individual observations into a single $\La_{\rm{obs}}^n$:

\begin{align}
\La_{\rm{obs}}^n = \prod_i^{\#{\rm{observations}}} \Lb_{{\rm{obs,i}}}^n.
 \end{align}

Having the likelihood of the observed dataset we find the probability of drawing this dataset from our simulated sample. We first make a PDF of the likelihood of each simulated optical depth value using the constructed KDEs. Then we use the $p$-value of $\La_{\rm{obs}}^n$ for this PDF as the probability of drawing the full observational dataset using the simulated model. Since the KDE is normalized this value is always between $0$ and $1$. An illustration of this procedure is shown in Figure~\ref{fig:likelihood}. We will discuss the result of our likelihood measurements in Section~\ref{deltaT-constraint} and further Figures showing the results of the inference are provided in Appendix~\ref{appendix-inference}.


\section{Results} \label{results}

\begin{figure}
    \centering
    \includegraphics[width = \linewidth]{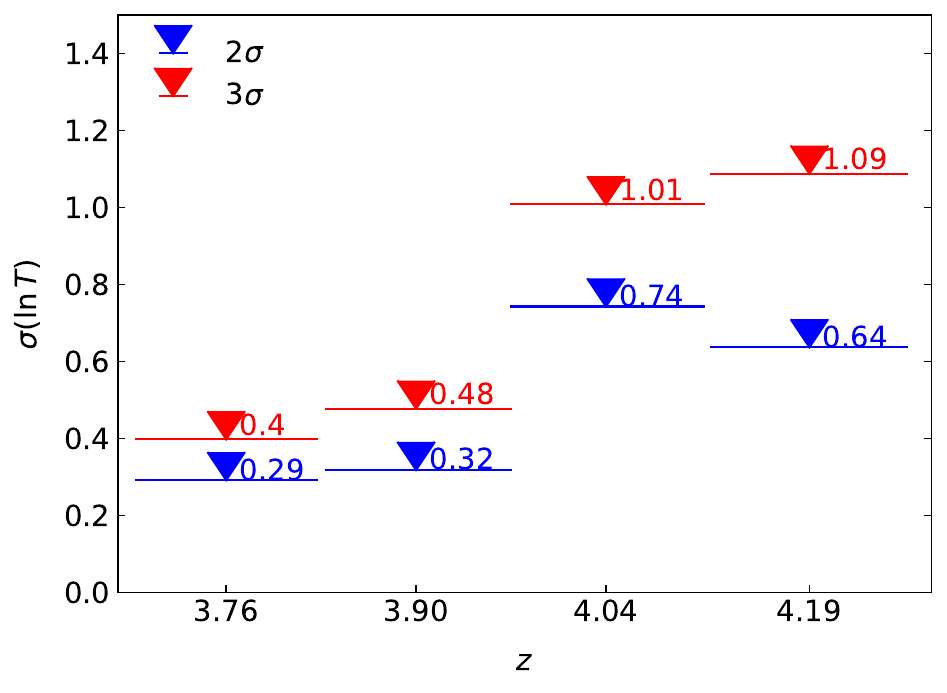}
    \caption{Summary of our constraints. The blue triangles show the $2\sigma$ limits while the red triangles are the $3\sigma$ limits. Our constraining power falls dramatically in the last two redshift bins due to the small number of sightlines. We constrain the amount of temperature fluctuations to be $\sigma(\ln T) < 0.29 \ (0.4) $ at $2\sigma\ (3\sigma) $, respectively. Constraints are shown for the $L=100\ \rm{Mpc}$ scale.}
    \label{fig:6money}
\end{figure}

\begin{table}
    \centering
    \begin{tabular}{c|c|c|c|c}
    \cline{1-5}
       $z$ & $2\sigma$, $100 \rm{Mpc}$  & $2\sigma$, $50 \rm{Mpc}$ & $3\sigma$, $100 \rm{Mpc}$  & $3\sigma$, $50 \rm{Mpc}$ \\
       \cline{1-5}
       3.76 & $<$0.29 & $<$0.40 & $<$0.40 & $<$0.58\\
       3.90 & $<$0.32 & $<$0.40 & $<$0.48 & $<$0.61\\
       4.04 & $<$0.74 & $<$0.88 & $<$1.01 & $<$1.19\\
       4.19 & $<$0.64 & $<$0.98 & $<$1.09 & $<$1.75\\
       \cline{1-5}
    \end{tabular}
    \caption{Summary of the $2\sigma$ and $3\sigma$ limits we obtain on the magnitude of temperature fluctuations $\sigma(\ln T)$ for two different physical scales.}
    \label{tab:constraints}
\end{table}

\subsection{Cumulative distribution functions of optical depths including temperature fluctuations} \label{CDF}

To illustrate the broadening of the effective optical depth distribution resulting from excess temperature fluctuations we plot the CDFs of effective optical depth in each redshift bin for both simulations and observations. Figure \ref{2-CDF} shows the CDFs resulting from our observations and Nyx without any excess temperature fluctuations (blue) and with $\sigma(\ln T)=0.5$ excess fluctuations (red). Our measurements in all redshift bins are in agreement with the simulations without any temperature fluctuations within $2\sigma$. Consequently, we explore the level of temperature fluctuations when the observations start to disagree with simulations at more than 2$\sigma$ and 3$\sigma$, this way we constrain the permitted $\sigma(\ln T)$. To produce quantitative constraints, we use the likelihood procedure as outlined above.

\subsection{Constraints on temperature fluctuations} \label{deltaT-constraint}
We use the likelihood approach described in Section~\ref{likelihood} to constrain the amount of temperature fluctuations which could be present. Figure \ref{fig:5pvalue} shows the $p$-value of occurrence of each of the Nyx+$\sigma (\ln T)$ models with different $\sigma (\ln T)$ values given our observations. Our constraining power drops in the last two redshift bins $z= 4.04$ and $4.19$ due to the relatively small number of sightlines in these two bins $23\ (24)$ and $13\ (13)$ respectively for the $100\ (50)\ \rm{Mpc}$ scale.
Comparing the results from two different scales demonstrates that we are more sensitive to temperature fluctuations at larger scales of $100\ \rm{Mpc}$ in comparison with $50\ \rm{Mpc}$, which is expected since the scatter in $\tau_{\rm{eff}}$ from density field fluctuations is smaller at larger scales, while the $\tau_{\rm{eff}}$ change arising from reionization-related temperature fluctuations depends on the topology of the reionization process, which may still be coherent on large scales.
Our strongest constraints are in the redshift bins $z=3.76$ and $z=3.90$, where we can constrain the amount of temperature fluctuations to be $\sigma(\ln T) < 0.29 \ (0.4) $ at $2\sigma\ (3\sigma)$. These limits correspond roughly to temperature contrasts between ionized and neutral regions of $\Delta T/T \sim 34\% \ (49\%)$.

For the last two bins, $z=4.04$ and $z=4.19$, the maximum likelihood model has non-zero temperature fluctuations. However, this preference is not statistically significant, since the model without any temperature fluctuations is still permitted within $2\sigma$. 
Larger quasar samples at higher redshifts are required to confirm this tentative detection of temperature fluctuations at $z>4$. The constraints we derived for all redshifts and physical scales are given in Table~\ref{tab:constraints}.

\section{Discussion}\label{discussion}



Using the Bayesian likelihood procedure described above, in our most sensitive redshift bin at $z=3.76$ we constrain the large-scale temperature fluctuations in the IGM to be less than $\sigma(\ln T)=0.29 \ (0.40)$ on 100\,Mpc (50\,Mpc) scales. This is comparable to the constraining power suggested by analyses of the large-scale Ly$\alpha$ forest power spectrum \citep{lya-power2005,Lai_2006}, but derived from a simple summary statistic of a relatively small number of high-quality quasar spectra. Our measurements are the only such constraints on temperature fluctuations from helium reionization thus far. Figure~\ref{fig:6money} summarizes our $2\sigma$ and $3\sigma$ constraints on $\sigma(\ln T)$ on 100\,Mpc scales as a function of redshift.

We note that our constraints neglect several features of helium reionization heating. As mentioned above, we treat the temperature increase as uniform across the given Ly$\alpha$ forest sightline, whereas in reality the fluctuations are unlikely to be so solidly coherent. In addition, we only modify the \emph{mean} temperature of the IGM, but it should also have an impact on the relationship between temperature and density. That is, the temperature-density relation $T=T_0 \Delta^{\gamma-1}$, where $\Delta$ is the baryon density relative to the cosmic mean and $T_0$ is the temperature at mean density, is typically shifted to lower values of $\gamma$ as the heat injection is not density-dependent. According to the $\tau_{\rm eff}$ scaling relation from \citet{bolton_et_al_2005}, incorporating $\gamma$ fluctuations correlated with $T$ fluctuations would slightly reduce their impact, but the maximum contrast in $\gamma$ is much smaller than that of $T$, so we do not expect this to have substantial implications for our analysis.

Few simulation predictions exist for the expected strength of large-scale temperature fluctuations from helium reionization. The most recent estimate comes from \citet{McQuinn_2009}, who performed radiative transfer post-processing of large-volume cosmological N-body simulations. In their largest volume simulations (429\,Mpc on a side), they found that the temperature fluctuations reached a peak of $\Delta T/T\sim0.2$ on 50-150\,Mpc scales very early on in the helium reionization process when the \ion{He}{3} fraction was $10\%$. The preference for excess $\tau_{\rm eff}$ fluctuations that we observe in the $z=4.04$ bin is consistent with this level of temperature fluctuations, and is thus consistent (but only suggestive) of this stage of the process.



\section{Conclusion}\label{conslusion}
In this work, we used the XQ-100 quasar sample \citep{xshooter2016} to search for excess temperature fluctuations in the IGM resulting from helium reionization. We began by reconstructing the underlying quasar spectra using the PCA method of \citet{bosman_2021}. 
We then measured the distributions of the effective optical depth of the Lyman-$\alpha$ forest towards the quasars and compared them with models of the IGM with an increasing amount of temperature fluctuations. 

We obtained constraints on the amount of excess $\sigma (\ln T)$ at two different spatial scales, $100\  \rm{Mpc}$ and $50\  \rm{Mpc}$, and four different redshifts: $z=3.76$, $z=3.90$, $z=4.04$ and $z=4.19$. We rule out temperature fluctuations as small as $\sigma (\ln T) = 0.29\ (0.40)$ at $2\sigma\ (3\sigma)$, with our tightest constraints being for scales of $100$ Mpc at $z=3.76$. Our measurements are the only such constraints to date. The constraining power of our new approach is comparable to forecasts from previous methods relying on the power spectrum of the Lyman-$\alpha$ forest. At $z=4.04$ and $z=4.19$, the observations modestly favor the presence of temperature fluctuations of about $\sigma (\ln T) = 0.4$; however, the detection is not statistically significant. 

We find that the distribution of effective optical depths has considerable constraining power on temperature fluctuations during helium reionization, with upper limits approaching the level predicted from cosmological radiative transfer simulations. Tighter constraints will require much larger samples of quasars -- assuming that the constraining power scales roughly as $1/\sqrt{N_{\rm qso}}$, consistent with our limits, suggests that $\sim400$ ($\sim700$) quasar sightlines of similar quality would be required to reach a $2\sigma$ sensitivity of $\sigma(\ln T)=0.1$ on 100\,Mpc (50\,Mpc) scales, which would then be sensitive to temperature fluctuations at the level predicted by \citet{McQuinn_2009} during most of helium reionization.

While we have focused on deep X-Shooter spectroscopy here to operate at high signal-to-noise and optimize our ability to predict the quasar continuum within the Ly$\alpha$ forest, more stringent constraints may be possible with existing and upcoming quasar spectroscopic samples from SDSS/BOSS, DESI \citep{DESI}, WEAVE-QSO \citep{WEAVE}, and 4MOST \citep{4MOST}, which have a few orders of magnitude more quasar sightlines at the cost of lower signal-to-noise and less coverage of the red-side quasar continuum. Future exploration of the effective optical depth distribution in these datasets may finally detect the temperature fluctuations from helium reionization.

We note that, while we have focused on the particular astrophysics of helium reionization, our methodology would be sensitive to any physical process that increases large-scale fluctuations in any quantity that modifies the opacity of the Ly$\alpha$ forest, i.e.~the UV background radiation (e.g.~\citealt{Pontzen14}) or even the underlying matter distribution. The agreement we find between the observations and simulations is thus representative of the success of the standard cosmological model.




\section*{Acknowledgments}
SER and SEIB are supported by the Deutsche Forschungsgemeinschaft (DFG) under Emmy Noether grant number BO 5771/1-1. SER is grateful for support from the Student Summer Internship program at the Max Planck Institute for Astronomy, which enabled this project to start. Based on observations made with ESO Telescopes at the La Silla Observatory under program ID 189.A-0424(A).

%



\bibliography{./Biblio.bib}{}
\bibliographystyle{aasjournal}

\appendix

\section{Bin selection} \label{appendix-bin}
In this Appendix, we detail our procedure for defining our redshift bins.  
Our bin selection procedure was initially optimized to use the entire Lyman-$\alpha$ forest spectrum of our accessible quasar sample spanning a range between redshifts $z=2.8$ to $z=4.4$. Starting from redshift $z=2.8$, we binned our data uniformly in consecutive chunks with comoving length of $100 \, \mathrm{Mpc}$. However, as later discovered, the spectra are affected by the fluxing error in the UV arm of the X-Shooter spectrograph, resulting in unreliable flux calibration between the X-Shooter's UV and VIS arms. After we attempted to stitch the two arms of the spectrum, this issue led to inconsistent flux levels in the two sides of the spectrum in the VIS and UV arms. The stitching occurs at a wavelength corresponding to roughly redshift $z=3.6$. 

\begin{figure*}
\centering
 \includegraphics[width=0.8\linewidth]{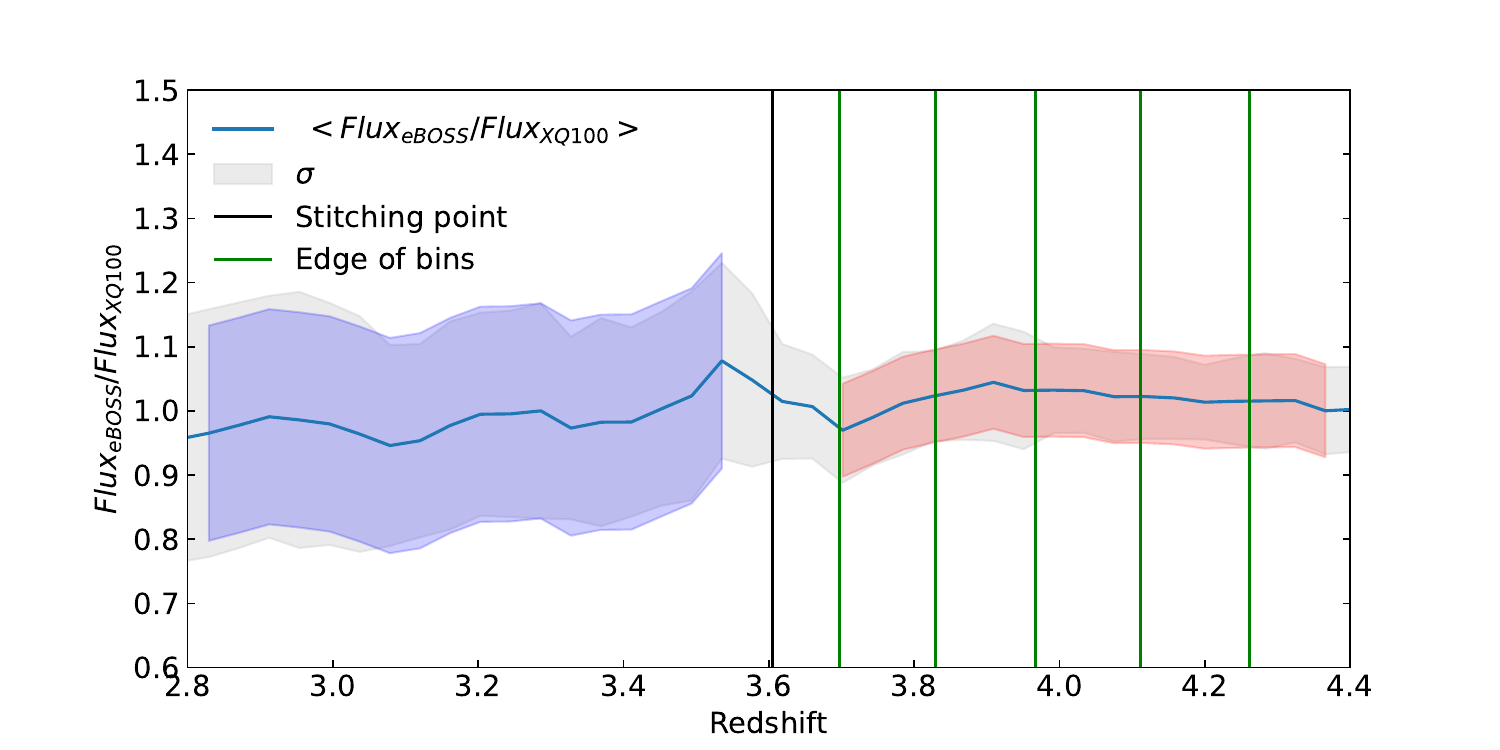}
    \caption{Flux ratio of spectra of the same quasars observed as part of the XQ-100 and eBOSS samples. As can be seen, the relative difference between the two increases by $>10\%$ after the stitching point, shown by a black vertical line, between the UV and VIS arms of the X-Shooter instrument. We use only the redshift bins delineated with green lines. The blue shaded region shows the average relative difference between XQ-100 and eBOSS in the UV arm of the X-Shooter instrument, while the red shaded region shows the difference in the VIS arm.}
    \label{fig:appx1}
\end{figure*}

Figure \ref{fig:appx1} shows a comparison between the fluxes of the same quasars observed as part of the XQ-100 and eBOSS samples. As can be seen, the relative error in the flux of the XQ-100 sample compared to the eBOSS sample increases by $>10\%$ after the stitching point at redshift $3.6$, with the flux scaling issue starting around redshift $3.68$. For this reason, we discarded all of our redshift bins at $z < 3.68$ and retained only the higher redshifts, with centers determined using our previous scheme. This results in $4$ redshift bins, each with a size of $100 \, \mathrm{Mpc}$, centered at redshifts $z=3.76$, $z=3.90$, $z=4.04$, and $z=4.19$.

\section{Likelihoods}\label{appendix-inference}
This Appendix provides additional plots related to our statistical inference. The following plots show the PDF of the likelihood of the simulated datasets compared to the likelihood of the observations at each redshift and for each value of $\sigma (\ln T)$. 
The plots are provided for the two spatial scales under consideration, $50$ Mpc and $100$ Mpc.

\begin{figure*}
\centering
 \includegraphics[scale=0.85]{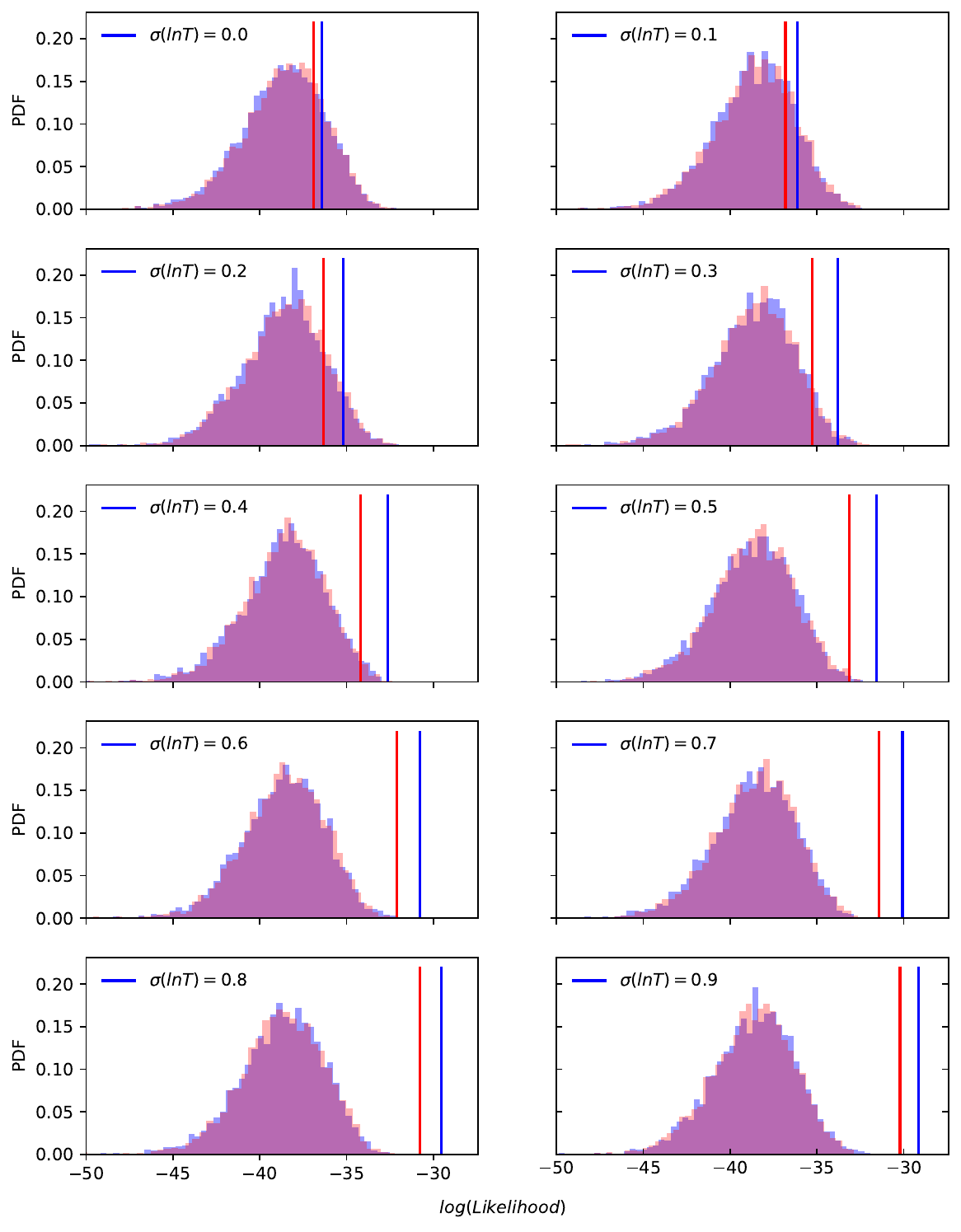}
    \caption{Likelihoods for the redshift bin $z = 3.76$, red corresponds to $L= 50 \rm{Mpc}$ and blue corresponds to $L= 100 \rm{Mpc}$. The histograms show the simulated KDEs and the solid line specifies the likelihood of our observations given this KDE.}
    \label{fig:like1}
\end{figure*}

\begin{figure*}
\centering
 \includegraphics[scale=0.85]{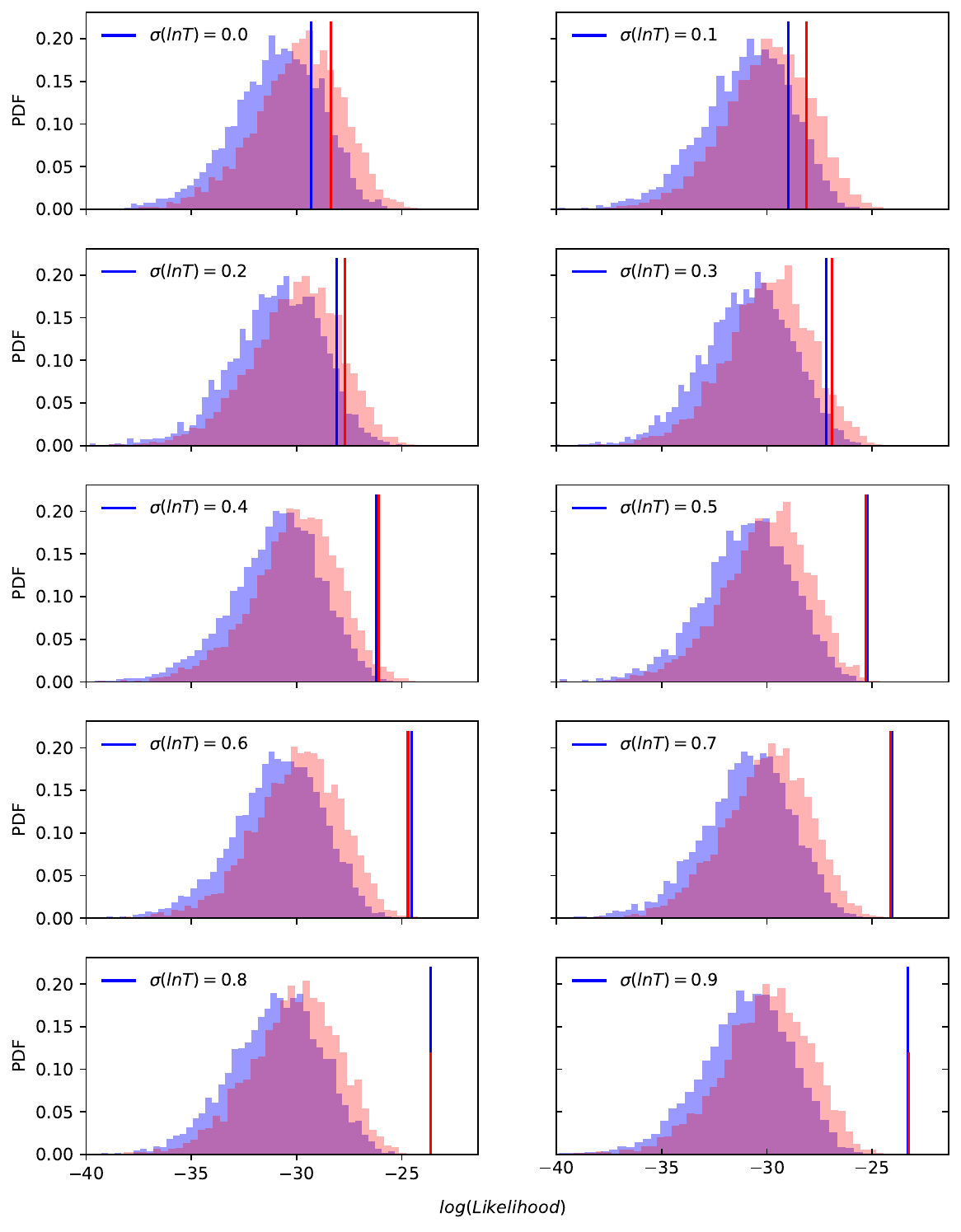}
    \caption{Likelihoods for the redshift bin $z = 3.90$, red corresponds to $L= 50 \rm{Mpc}$ and blue corresponds to $L= 100 \rm{Mpc}$. The histograms show the simulated KDEs and the solid line specifies the likelihood of our observations given this KDE.}
    \label{fig:like2}
\end{figure*}

\begin{figure*}
\centering
 \includegraphics[scale=0.85]{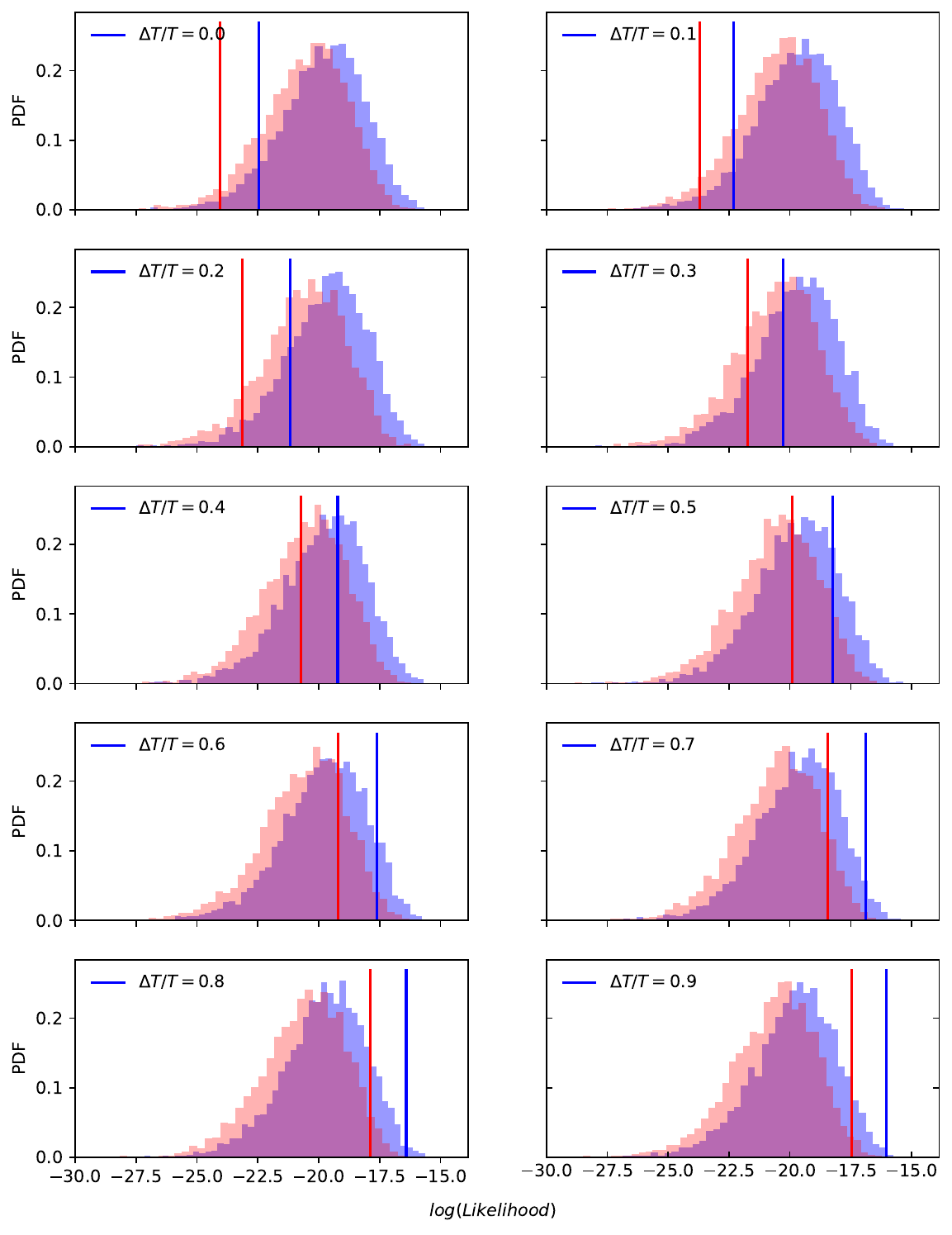}
    \caption{Likelihoods for the redshift bin $z = 4.04$, red corresponds to $L= 50 \rm{Mpc}$ and blue corresponds to $L= 100 \rm{Mpc}$. The histograms show the simulated KDEs and the solid line specifies the likelihood of our observations given this KDE.}
    \label{fig:like3}
\end{figure*}

\begin{figure*}
\centering
 \includegraphics[scale=0.85]{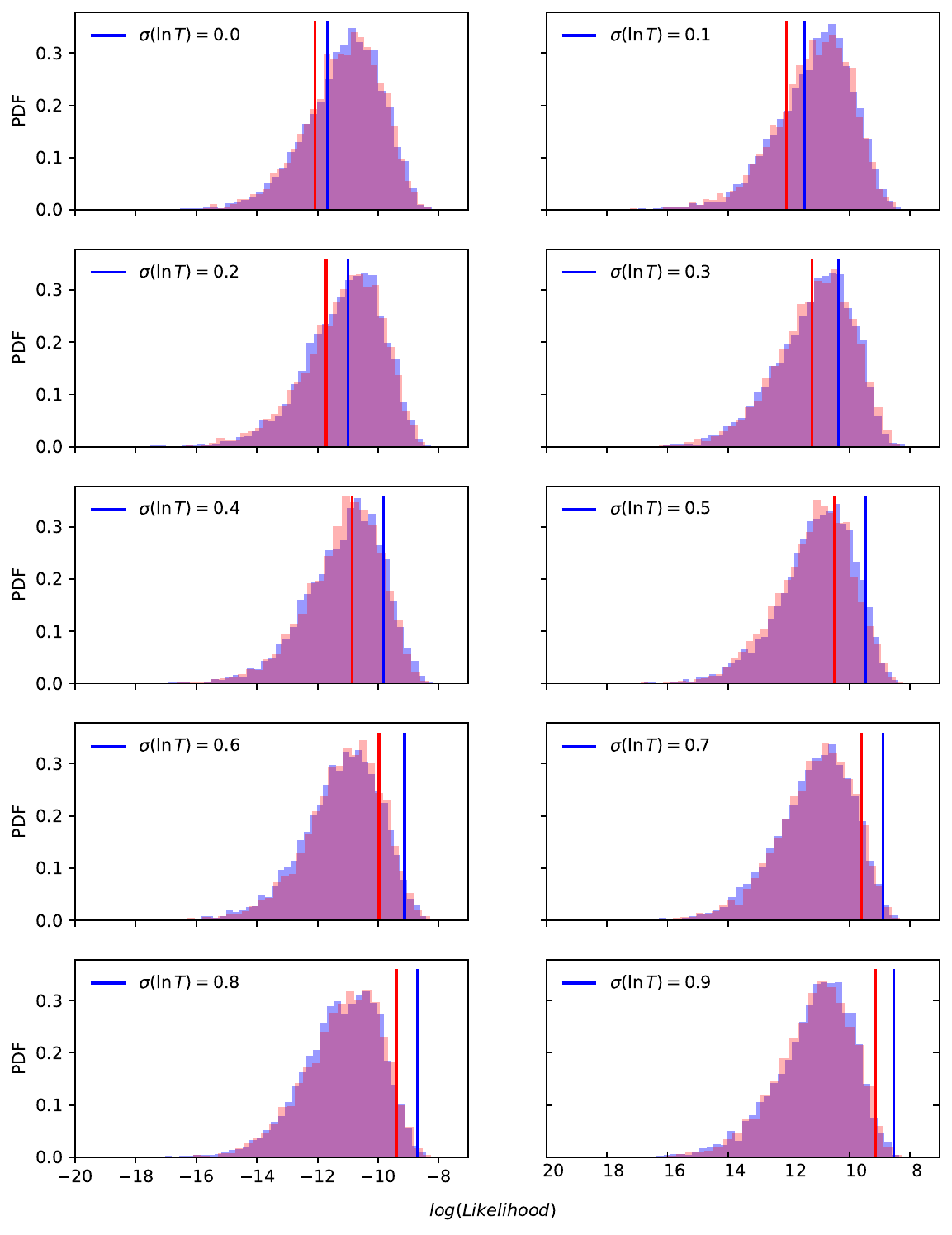}
    \caption{Likelihoods for the redshift bin $z = 4.19$, red corresponds to $L= 50 \rm{Mpc}$ and blue corresponds to $L= 100 \rm{Mpc}$. The histograms show the simulated KDEs and the solid line specifies the likelihood of our observations given this KDE.}
    \label{fig:like4}
\end{figure*}



\end{document}